\documentclass[journal]{IEEEtran}

\usepackage{amsmath}
\usepackage{amssymb,graphicx}

\usepackage[noadjust]{cite} 

\usepackage{color}
\usepackage{ulem}

\newcommand{\be}{\begin{equation}}
\newcommand{\bea}{\begin{eqnarray}}
\newcommand{\ee}{\end{equation}}
\newcommand{\eea}{\end{eqnarray}}

\newcommand{\changed}[1]{#1}

\newcommand{\includefigure}[1]{\centerline{\includegraphics[width=\columnwidth]{#1}}}
\def\s{\mbox{$s$}}
\def\n{\mbox{$n$}}
\def\g{\mbox{$g$}}

\def\J{\mbox{$\cal{J}$}}
\def\JM{\mbox{$\cal{J}_M$}}

\def\gs{\mbox{$\gamma_s$}}
\def\gc{\mbox{$\gamma_c$}}
\def\gp{\mbox{$\gamma_p$}}
\def\gn{\mbox{$\gamma_n$}}

\begin{document}

\title{Shaping Current Waveforms for direct Modulation of Semiconductor Lasers}
%
%
\author{%
Lucas~Illing and Matthew~B.~Kennel%
\thanks{
\changed{%
This work was supported in part by the US Army Research Office (grant \# DAAD19-99-1-0199 and grant \# DAAG55-98-1-0269)}
}%
\thanks{L.~Illing is with the Department of Physics and the Center for Nonlinear and Complex Systems, Duke University, P.O. Box 90305, Durham, NC 27708 (e-mail:illing@phy.duke.edu)}%
\thanks{M.~B.~Kennel is with the Institute for Nonlinear Science, University of California, San Diego, La Jolla, CA 93093-0402}%
}

\maketitle

\begin{abstract}
We demonstrate a technique for shaping current inputs for the direct
modulation of a semiconductor laser for digital communication.
\changed{ The introduction of shaped current inputs allows for the
suppression of relaxation oscillations and the avoidance of
dynamical memory in the physical laser device, i.e., the output
will not be influenced by previously communicated information.}
On the example of time-optimized bits, the possible performance
enhancement for high data rate communications is shown
\changed{numerically.}
\end{abstract}


\IEEEpeerreviewmaketitle

\section{Introduction}

\PARstart{S}{emiconductor} lasers typically convey information for
optical communication using either ``external'' or ``direct''
modulation.  In the first scheme, the laser is maintained in a
constant light-emitting state and some sort of external ``shutter,''
typically a Mach-Zehnder type electro-optic modulator, modulates the
output intensity according to an externally applied voltage. The
advantage of this technique is that it allows high bit rates
\changed{($>10$ Gbit/s) } and there is no interaction between the
information being applied to the modulator and any internal dynamics
of the laser.  The disadvantage is that costly external apparatus are
necessary.  Direct modulation involves changing the current input of
the laser, i.e., modulating it around the bias current, in order to
produce time-dependent output in optical intensity.

\changed{In terms of modeling, the use of direct modulation means that
  the laser becomes a driven nonlinear system with potentially
  complicated dynamics. Indeed, it is well known that even simple
  high-frequency periodic modulation of sufficiently large amplitude
  can lead to nonlinear phenomena such as period doubling cascades,
  period tripling, and chaos
  ~\cite{dml-lee,dml-liu,dml-Tang,dml-agr,dml-Lauterborn,dml-Lamela,dml-corbett,dml-exp2}.
  It is therefore not surprising that attempting communication with
  current pulses that occur rapidly in succession, quickly changing
  among discrete current levels to transmit randomly interleaved zero
  and one bits, will result in a complex output time series of
  intensity.  As a result the bits may cease to be decodable at the
  receiver and there is a considerable memory
  effect~\cite{colet_mirasso}. This effect is called ``inter-symbol
  interference'' in the communications literature because the sequence
  of information symbols at previous times may harm decoding at the
  present time.  Additionally, communication using high-speed direct
  modulation is limited by transient oscillations, known as relaxation
  oscillations, that result from the interplay between the optical
  field and the carrier density~\cite{Petermann,Lau}. Such
  oscillations cause distortions of the output light pulse and broaden
  the signal's optical spectrum (increased chirp), thus contributing
  to an increased bit-error rate.  }

\changed{In an effort to overcome the problems associated with direct
  modulation many techniques for chirp reduction and suppression of
  relaxation oscillations have been proposed, including the
  modification of the physical laser
  device~\cite{Lau,Tucker,Riyopoulos}, the use of external optical
  feedback~\cite{Mirasso} and external electronic
  circuits~\cite{Suematsu}, as well as the shaping of the current
  pulses that are used to encode logical bits into the optical output
  of the laser~\cite{Danielsen,Torphammar,Olshansky,Bickers,Dokhane}.
}

We approach the issue \changed{ of inter-symbol interference
  and relaxation oscillations } from the point of view of nonlinear
dynamics. A basic single-mode semiconductor laser modeled by the rate
equations is a two degree of freedom system, with ordinary
differential equations for photon density and carrier density. The
cause of inter-symbol interference is the fact that even if the
observable intensity returns to its nominal stationary condition after
a current wave form is applied, the unobservable carrier density may
not have, and thus there is dynamical coupling from past to future.
Our feed-forward solution is to use appropriately shaped driving
currents to control the state-space trajectory of the laser, and we
show how to obtain these waveforms given a dynamical model for the
laser.

\changed{The main goal of this paper is to demonstrate a technique
  that can eliminate relaxation oscillations and inter-symbol
  interference, without having to physically change the laser diode.
  This allows enhanced data rates and improved communication
  performance as is shown numerically for a communication scheme using
  time-optimized non-return-to-zero (NRZ) bits. This improvement comes
  at the cost of more complex input electronics, which would be
  necessary to shape the input current in experiments.  A
  distinguishing feature of our approach is the use of shaped,
  continuous, drive currents with finite rise times. This contrasts
  with other recent research~\cite{Dokhane}, where relaxation
  oscillations are suppressed using piece-wise constant currents,
  which would possibly present difficult discontinuities.}

We will first present the general idea of the shaping method \changed{
  in Sec.~\ref{current_shaping} } and then show particular solutions
for the specific case of optimizing the transition time for NRZ bits
in \changed{ Sec.~\ref{communication}.} Using these solutions, we
\changed{estimate} the bit error rate (BER) performance for the case of a
simple Gaussian white noise channel to show the possible performance
enhancement.

\section{Current Shaping}
\label{current_shaping}
The semiconductor laser is presumed to be single mode. For this case
the laser dynamics can be described by a driven two dimensional system
of ordinary differential equations for the carrier density $N(t)$ and
the photon density $S(t)$, the latter being proportional to output
intensity:
\begin{equation}
\begin{split}
\frac{dS}{dt} &= -\gc \, S + \Gamma \, G(N,S) \, S,  \\
\frac{dN}{dt} &= -\gs \, N  -G(N,S) \, S + \frac{J_0 + J}{ed}.
\end{split}
\label{rate}
\end{equation}
The constant $J_0$ is the bias current, $J(t)$ the modulation around the
bias, $G(N(t),S(t))$ the optical gain coefficient including nonlinear
effects, $\Gamma$ the confinement factor, $d$ the active layer
thickness of the laser, $\gc$ the photon decay rate, and $\gs$ the
spontaneous carrier decay rate.

For the purpose of numerical calculations and analytical
investigations it is useful to work with dimensionless quantities.  It
is convenient to non-dimensionalize using the fixed point $(S_0, N_0)$
of the laser in the absence of modulation ($J(t)=0$), that is, of the
laser under stationary continuous-wave operating conditions
$(dS/dt=dN/dt=0)$.  We expand the nonlinear $G(N(t),S(t))$ around this
point to first order \be G(N,S) = G_0 + G_n (N-N_0) + G_p
(S-S_0).\label{gain:taylor} \ee Liu and
Simpson~\cite{article:liusimpson} show how to experimentally estimate
$G_n =\partial G / \partial N >0$ and $G_p= \partial G/ \partial S<0$.
Note that the fixed point conditions yield the relations $\Gamma G_0 =
\gc$ and $J_0/ed -\gs N_0 = G_0 S_0$.

These dimensionless dynamical equations are independent of how far
above threshold the laser is operated except via indirect influence on
the empirical differential gain parameters $G_n$ and $G_p$, which
depend on the expansion point.  The dimensionless photon density
$\s(t)$ is defined via $\s(t)=S(t)/S_0$, the carrier density $\n(t)$
via $\n(t)= N(t)/N_0-1$, and the current via $\J + \JM =
(J_0+J(t))/(\gs N_0 e d) - 1$.%
\footnote{%
  \changed{ Roughly speaking, the dimensionless total current
    $1 + \J + \JM$ is in units of the threshold current. This is
    strictly true for simple gain functions that imply clamping of the
    carrier density to it's threshold value for currents above
    threshold.  }  } The dimensionless gain $\g=G/G_0$ is \be
 \label{gain:nondim}
 \g(\n,\s) = 1+\frac{\gn}{\J \gs} \n - \frac{\gp}{\gc} (\s-1),
\ee\index{gain}
where $\gn = G_n S_0$ and $\gp = -\Gamma G_p S_0$.
Using the natural relaxation oscillation angular-frequency $\omega_R =
\sqrt{\gc \gn + \gs \gp }$, the dimensionless quantity corresponding to
time $\tau = t \omega_R$ can be formed. With this the model equations
are:
%
\be
\begin{split}
 \frac{d\s}{d\tau} &=  \frac{\gc}{\omega_R} \Bigl( \bigl[\g(\n,\s)-1\bigr]\; \s \Bigr)  \\
 \frac{d\n}{d\tau} &= \frac{\gs}{\omega_R} \Bigl( \J + \JM - \n -  \J \;\g(\n,\s)\; \s \Bigr) . \\
\end{split}
\label{rate2}
\ee
Table~\ref{table:nondimrate}  lists the dimensionless quantities and
the parameters  we use, the values of  which are based on measurements
of J.M. Liu's group~\cite{article:Abarbanel}.
%
%
\begin{table}
\renewcommand{\arraystretch}{1.3}
\caption[Semiconductor laser variables and parameters]{Dynamical Variables, definitions, and numerical values of parameters.}
\label{table:nondimrate}
\begin{center}
\begin{tabular}{c|c|l}
\hline
Symbol&Value&Description \\
\hline
$\s(\tau)$& &dimensionless photon density\\
$\n(\tau)$& &dimensionless carrier density\\
$\g(\n,\s)$& &dimensionless gain coefficient\\
$\JM(\tau)$& &dimensionless modulation current\\
\hline
\J & 2/3& bias current at fixed point \\
 $\gs$ & 1.458 $\times 10^9 s^{-1}$ & spontaneous carrier decay rate \\
$\gn / \J$ & 2.0  $\times 10^9 s^{-1}$ & gain variation with carrier density\\
$\gp / \J$ & 3.6 $\times 10^9 s^{-1}$ & gain variation with photon density \\
$\gc$ & 3.6 $\times 10^{11} s^{-1}$ & photon decay rate \\
\changed{$\omega_R$} & $\sqrt{\gc \gn + \gs \gp} \sim \frac{1}{45 \, \text{ps}}$ & relaxation oscillation frequency\\
\hline
\end{tabular}
\end{center}

\end{table}
For the purpose of digital communication we are interested in the
relationship between the information bearing modulation current and
the output of the laser. The key insight is to note that, given a
functional form for the time series $s(t)$, the corresponding input
current can be computed just by differentiation.  In contrast, given
the functional form of the current, it is necessary to integrate the
nonlinear equations of motion (\ref{rate2}) to obtain the potentially
chaotic resulting laser output.

Given a particular model for the laser, there is often some
differential operator taking a given $s(\tau)$ to the other variables
$O_s[s(\tau)] = \JM(\tau)$ and another one $O_n[s(\tau)] = n(\tau)$
such that the state vector $[\JM(\tau), n(\tau), s(\tau)]$ is a
solution to the equations of motion.  This solution though may not be
physically allowable (such as having negative total current) or
dynamically stable.

To derive the driving current as function of the output waveform we
introduce, for computational ease, $\s(\tau)=e^{y(\tau)}$, solve for
$n(\tau)$ using the $ds/d\tau$ equation, differentiate, and rearrange
the $dn/d\tau$ equation to solve for $\JM(\tau)$, yielding
%
\bea \JM = \frac{\omega_R^2 \J}{\gamma_c
  \gamma_n} \left[ \ddot y + \left( \frac{\gamma_p
      +\gamma_n}{\omega_R} \;e^y + \frac{\gamma_s}{\omega_R} \right)
  \dot y + (e^y-1) \right].
\label{JM}
\eea
This equation relates a given waveform $s(\tau)$, via $y$, to the
modulation current that caused it.  Any desired output waveform that
corresponds to a physically allowed input can thus be produced by
driving the laser with the according shaped current $\JM(\tau)$. Not
all functions $\s(\tau)$ will however result in realistic $\JM(\tau)$.
One physical requirement is a non-negative total driving current at
all times, $1+\J+\JM \ge 0$, as it is not possible to extract
electrical energy back from the \changed{ carrier density.  }
Practical considerations generally limit the peak current that can be
applied as well.

Note that (\ref{JM}) includes the full nonlinear dynamics of the laser
as modeled by the rate equations (\ref{rate}).  In contrast to any
theory based on the linearization of (\ref{rate}), results obtained
using (\ref{JM}) apply to large driving amplitudes.  The specific
linearized parameterization of gain we used (\ref{gain:taylor})
restricts the validity of (\ref{rate2}) because it does not take into
account gain saturation effects that become important for very large
intracavity photon densities $s$.  However, it is equally
straightforward to derive formulas analogous to (\ref{JM}) when gain
functions which model saturation are used.  We chose to work with gain
(\ref{gain:taylor}) because it is simple.  For the results presented
in the rest of the paper gain saturation is not important, and, the
parameters of (\ref{gain:taylor}) may be estimated experimentally.

\section{Application to Communication}
\label{communication}

We consider a ``direct'' modulation setup: the laser's driving current
$J(t)$ is modulated to encode information.  The laser light travels
through the communication channel, and at the receiver, the intensity
of the arriving signal is measured. One then attempts to extract the
information from this signal, in the simplest case by a single
thresholding to decode binary information.

%
%
%
\begin{figure}
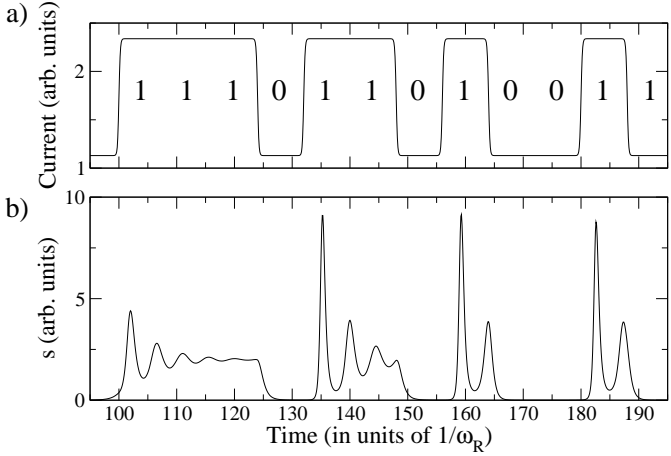

\includefigure{fig1_600dpi.eps}

\caption[Pulse shaping: relaxation oscillations]{We show how a square-like total driving current 
  \changed{ $1+\J+\JM$ } (a) gives rise to relaxation oscillations of
  the laser photon density $s(t)$ (b). The bit time is $T_{bit}=8$
  \changed{ which corresponds to 360 ps for the model-parameters shown
    in Tab.~\ref{table:nondimrate} ($\omega_R^{-1} \sim 45$ ps).  }
  The lower and upper value for the driving current corresponds to
  fixed points with $s=0.2$ and $s=2.0$, respectively.}
\label{fig:relaxation_osc}
\end{figure}

Figure~\ref{fig:relaxation_osc} shows the laser dynamics driven with
an unshaped current waveform (NRZ bits).  After a transition from low
current to high current, there are clear relaxation oscillations of
the laser output, that is, the output ``rings'' at the characteristic
relaxation oscillation frequency. In the particular situation here,
the bit time would be sufficiently large that a conventional system
could decode the information successfully by integrating energy over a
bit interval. A further increase in bit rate, however, will result in
significant inter-symbol interference (see Fig.~\ref{fig:shpvssqr}d-f)

The state-space trajectory depends on its initial conditions $s^0$ and
$n^0$ as well as on the driving signal $\JM(\tau)$. If, at the end of
each bit time $T_{bit}$, the system returns to a particular point in
state-space, e.g. $s(T_{bit})=s^0$ and $n(T_{bit})=n^0$, then
repeating the same driving signal $\JM(\tau)$ will result in a
repetition of the state-space trajectory.  If, on the other hand, the
system trajectory did not reach ($s^0,n^0$) after the first bit, then
repeating the driving signal will result in a different laser output,
whose particular details depend on the previous drive signal. This
dynamic coupling causes inter-symbol interference.

The main physical issue is to return the unobserved \changed{ carrier
  density } to a specific state as well as the photon density.  We use
(\ref{JM}) to design driving signals which return the system
trajectory to a known state at the end of each bit.  Since the adverse
effects of dynamic coupling on communication performance are
especially relevant for high bit rates, we will try to maximize the
bit rate (minimizing the bit time) for a given physical laser by
optimizing driving signals which cause no dynamical inter-symbol
interference.

\subsection{Finding the time-optimal NRZ bits}

NRZ bits encode 1-bits (0-bits) as high (low) light power, but in
contrast to return-to-zero pulsing, the intensity stays high when two
or more consecutive 1-bits are transmitted, and similarly for
consecutive zero bits.  The dynamical picture is a system that can be
driven to switch between two fixed points in state-space, one
corresponding to low output power, the ``off'' state, and the other
corresponding to high output power, the ``on'' state.  The driving
waveform for a bit identical to its predecessor is trivial---a
constant to stay at the particular fixed point.  We solve for driving
currents that fix the system along a state-space trajectory, smoothly
connecting the ``off'' fixed point to the ``on'' fixed point (up
transition) and another one for the ``on'' to ``off'' transition (down
transition). We require both the photon density and the carrier
density come to a known fixed point at the end of each bit, thus there
will be no dynamic coupling from one bit to the next and therefore no
inter-symbol interference.

We use (\ref{JM}), a parameterized set of photon density wave forms,
and an optimization algorithm to search for waveforms which optimize
some other desirable communication criterion under the condition that
they correctly connect the two states.  In particular, we have
searched for the minimum transition time $T$, and thus highest bit
rate, given a desired intensity separation between high and low.

To be specific, let us fix the ``off'' intensity to 20\% of the fixed
point intensity around which we normalized, $\s_{\text{off}}=0.2$, and
the ``on'' state to twice the fixed point value, $\s_{\text{on}}=2.0$.
For the up (down) transition we need to search the space of all smooth
waveforms $\s(\tau)$ with $\s(0)=\s_{\text{off}}$ and
$\s(T)=\s_{\text{on}}$ ($\s(0)=\s_{\text{on}}$ and
$\s(T)=\s_{\text{off}}$) such that the transition time $T$ is
minimized and such that the corresponding current $\JM(\tau)$ meets
certain constraints. In this example, we require that $0<1+\J+\JM<6$,
which means that the total input current into the laser is positive
and that the largest pump current we admit is roughly six times the
threshold current, a physically reasonable requirement.  We
furthermore require that $\JM(\tau)$ has zero first derivatives at
$\tau=0$ and $\tau=T$.

The success of the optimization over all possible waveforms depends on
the choice of the right set of basis functions, \changed{and
  satisfying all the boundary conditions}.  We found that, although an
ordinary Fourier base does work to a certain extent, it is \changed
{numerically advantageous to pick base functions which automatically
  incorporate many of the constraints, avoiding needing as many
  explicit constraint equations in the optimization procedure.  This
  notion led us to use the following functions to parameterize}
$s=e^y$: \be\label{base}
y(\tau)=p\left(\frac{\tau}{T}\right)+\left[\sum_{k=1}^N a_k
  \sin\left(2\pi f\left(\frac{\tau}{T}\right)\right) \right]^4 , \ee
where $p(\cdot)$ is a (7th degree) polynomial with
$p(0)=\ln\left[\s(0)\right]$, $p(1)=\ln\left[\s(T)\right]$, and with
zero first, second, and third derivative at $\tau=0$ and $\tau=T$,
which ensures zero first derivatives of $\JM$ at these times.
$f(x)=(a_0 +1)x / (a_0 x +1)$ is a warp function introducing asymmetry
with the effect that short transition times can be obtained using
fewer base functions.  The only constraint that has to be enforced in
the optimization is the requirement $0<1+\J+\JM<6$ (see
Table~\ref{table:optparams}).  \changed{This choice of base
functions was rather
  arbitrary, justified principally by numerical ease in automatically
  incorporating boundary conditions and empirically giving reasonably
  successful results, not by any strong theoretical considerations.
  We leave the question of how to choose mathematically optimal, or
  experimentally desirable, base functions for future investigations.
  Our main goal is to demonstrate the general concept; there is no
  barrier to incorporating other forms of base functions and other
  models for different lasers in this method. }

%
%
%

\begin{figure}
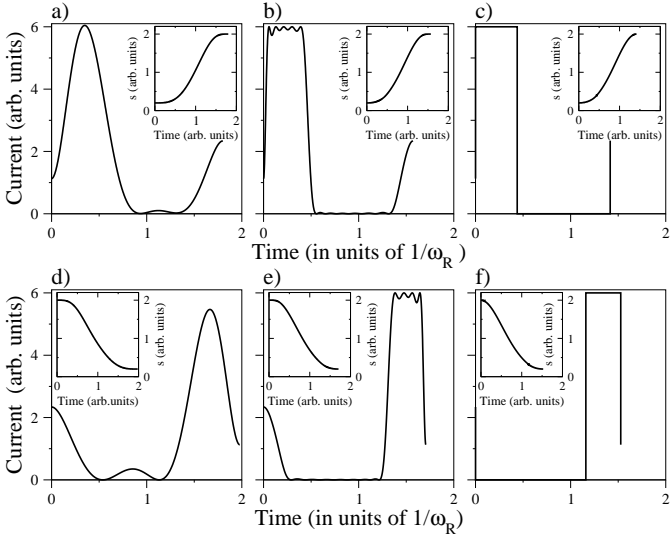

\includefigure{fig2_600dpi.eps}
\caption[]{ 
  \changed{ The figure shows driving current waveforms, i.e.
    $1+\J+\JM(t)$, and in the insets the corresponding optical laser
    output in terms of the photon density $s(t)$.  Panels a), b), and
    c) show the transition from the ``off'' to the ``on'' state. Using
    two sin-terms [N=2 in (\ref{base})] in the optimization procedure
    for the driving current yields a), whereas the use of 20 base
    functions (N=20) results in b). For comparison, we display in c)
    the photon density waveform and discontinuous current computed
    using optimal control theory.  Panels d), e), and f) are results
    for the ``on'' to ``off'' transition, where two base functions are
    used in d), 20 base functions in e), and f) is the optimal control
    theory result.  } }
\label{fig:uptopt}
\end{figure}

Figure~\ref{fig:uptopt} shows numerical results from using base
functions (\ref{base}) and the NPSOL software
package~\cite{NPSOL1,NPSOL2}, which allows nonlinear constrained
optimization.  \changed {Comparing the currents for the
  ``off'' to ``on'' transition (Fig.~\ref{fig:uptopt}a, b, and c) to
  the corresponding photon densities $s(t)$ (insets) shows that the
  shaped current waveform that is obtained as result of the
  optimization is quite dissimilar from the output intensity waveform.
}

In particular, an increase of the laser's output intensity is not
achieved by just ramping up the current from the low value
corresponding to the ``off'' steady state to the higher value
corresponding to the ``on'' state. Instead the pump current is reduced
after an initial surge to allow the carrier density to relax to the
steady state value.

In Fig.~\ref{fig:uptopt}a we show the current and photon density that
were obtained with just two sinusoidal terms (N=2 in (\ref{base})).
The minimal transition time that can be obtained decreases when the
number of base functions is increased, e.g. tenfold (N=20) as shown in
Fig.~\ref{fig:uptopt}b.  The inclusion of higher frequency sine
functions results in a current with a much faster rise time.  Just by
the choice of the number of base functions one can therefore
approximately take into account limitations on the possible rise-times
of electronics involved.  Thus, we did not explicitly add this
limitation as an additional constraint in the optimization procedure,
but such a constraint is reasonable to use.
 
The waveforms found also converge to the theoretical best limit
\changed{ (Fig.~\ref{fig:uptopt}c), } obtained by applying
results from optimal control theory. If one drops the requirement of
finite rise times for the drive current, that is, if discontinuous
functions are admitted, then the so called Minimum
Principle~\cite{book:Athans} can be applied. It tells us that a
time-optimal state-space trajectory is achieved by driving the laser
with a constant current at either the maximal allowed or minimal
possible value \changed { (see as well~\cite{Dokhane}).
} The point in state-space where to switch from one value to the other
can be found by integration of the Hamiltonian corresponding to
(\ref{rate2}) (for details see~\cite{mythesis}). The resulting minimum
transition time is a lower bound benchmark for the transition times
that can be achieved with realistic input currents. It also is the
lower bound for the bit period $T_{bit}=T$ for which communication
without laser induced inter-symbol interference is possible.  Figure
\ref{fig:uptopt} shows that the results we obtain with our technique
converge both in the shape of the waveform and in the value of the
transition time to the optimal-control theory result as more base
functions are included in the optimization routine.

%
%
%
%
%
\begin{table}
\renewcommand{\arraystretch}{1.3}
\caption{Summary of parameters, constraints, and results for the optimized NRZ bits}
\label{table:optparams}
\begin{center}
\begin{tabular}{|l|c|c|c|}
\hline
\multicolumn{4}{|c|}{ Fixed parameters in the optimization } \\
\hline
\multicolumn{2}{|l}{0-bit fixed point} 
&\multicolumn{2}{|c|}{$s_{0bit}=0.2$}  \\
\multicolumn{2}{|l}{1-bit fixed point} 
&\multicolumn{2}{|c|}{ $s_{1bit}=2.0$} \\
\multicolumn{2}{|l}{minimal allowed pump power }
&\multicolumn{2}{|c|}{ $1+\J+\JM^{min}=0.0$} \\
\multicolumn{2}{|l}{maximal allowed pump power }
&\multicolumn{2}{|c|}{$1+\J+\JM^{max}=6.0$ }\\
\hline 
\hline
\multicolumn{4}{|c|}{ Optimization constraints and how they are met} \\
\hline
\multicolumn{2}{|l}{fixed $s(0)$ (either $\s_{\text{on}}$ or $\s_{\text{off}}$)} 
&\multicolumn{2}{|c|}{$y(0)=p(0)=\ln(s(0))$}  \\
\multicolumn{2}{|l}{fixed $s(T)$ (either $\s_{\text{off}}$ or $\s_{\text{on}}$)} 
&\multicolumn{2}{|c|}{$y(T)=p(1)=\ln(s(T))$} \\
\multicolumn{2}{|l}{$\frac{\partial}{\partial t} \JM(0) =0$} 
&\multicolumn{2}{|c|}{$\dot y(0)=\ddot y(0)= \dddot y(0)=0$} \\
\multicolumn{2}{|l}{$\frac{\partial}{\partial t} \JM(T) =0$} 
&\multicolumn{2}{|c|}{$\dot y(T)=\ddot y(T)= \dddot y(T)=0$} \\
\multicolumn{2}{|l}{$0<1+\J+\JM(\tau)<6$} 
&\multicolumn{2}{|c|}{ active constraint} \\
\hline
\hline
\multicolumn{4}{|c|}{ Results of optimization} \\
\hline
     & N=2 & N=20 & Theory \\
\hline
 up transition   & $T=1.80$ & $T=1.57$ & $T_{opt}=1.42$ \\
 down transition & $T=1.98$ & $\quad T=1.71 \quad$ & $T_{opt}=1.53$ \\
\hline
\end{tabular}
\end{center}
\end{table} 
%
%

The down transition driving the laser from the ``on'' power level to
the ``off'' power level \changed{ (Fig.~\ref{fig:uptopt}d, e,
  and f) } we design exactly as before except that values used for the
boundary conditions $s(0)$ and $s(T)$ are reversed from the low to
high transition.  The minimal transition times are displayed in
table~\ref{table:optparams}, where we show the results using two and
twenty base functions as well as using optimal control theory.  It is
important to note that the down transition is not just a time reversed
version of the up transition, and that both the transition times and
the waveforms are different and have to be designed separately.  Most
communication schemes are evenly clocked, the time necessary to
transmit a symbol is independent of the particular information.  Yet
our results show that the fundamental dynamics of the laser may mean
that approaching the optimal limits might imply using unequal symbol
times.  We offer this as an idea to explore but we do not pursue it
further.

\subsection{Performance}

Using shaped input currents will result in a better BER performance
for high bit rates because we avoid errors caused by inter-symbol
interference due to the internal dynamics of the laser. To show the
possible improved communication performance we compare NRZ bits
obtained using the up and down transitions that were designed by the
presented technique using two base functions (N=2 in (\ref{base})) to
unshaped current modulation.

\begin{figure}
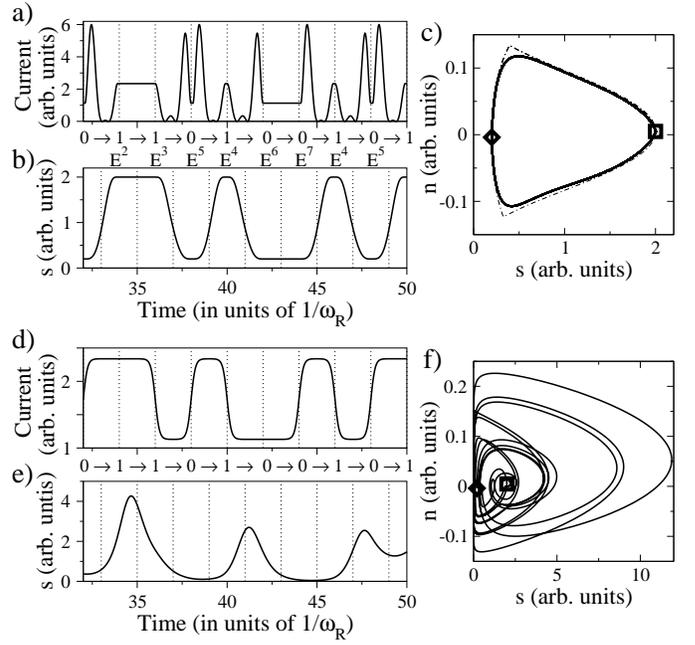

\includefigure{fig3_600dpi.eps}
\caption[Non-return-to-zero bits state-space plot]{
  \changed{ Comparison between shaped current modulation (a-c) and
    square-like currents (d-f).
Shaped drive currents (a) result in an optical laser output (b) free
of relaxation oscillations.  For visual guidance, bit times are
indicated by dotted lines.  Note that the currents were designed to
optimize the transition between the two logical states (``on'' and
``off'' ) and consequently, in the output, the assigned bits appear
shifted by half a bit time. The bit assignment is shown between panels
a) and b) together with the corresponding bit energies (see text).
Square-like currents (d) result in an output (e) corrupted by strong
inter-symbol interference.  Panels c) and f) are the corresponding
state-space plots (see text for further details).  } }
\label{fig:shpvssqr}
\end{figure}

A random sequence of bits is to be transmitted in this numerical
experiment. The bit sequence determines the shaped current, shown in
Fig.\ref{fig:shpvssqr}a, which enters as driving term the laser rate
equations (\ref{rate2}). The output $s(\tau)$ is obtained by
integration of (\ref{rate2}) and Fig.\ref{fig:shpvssqr}b shows that
$s(\tau)$ switches, as was the design goal, between the ``on'' and
``off'' state without any relaxation oscillations. Transmitting 50
bits results in the state-space trajectory presented in
Fig.\ref{fig:shpvssqr}c (thick line). The two fixed points (square and
diamond) and the optimal control trajectory (dash-dotted line)
obtained by using non-continuous currents are also shown.

The bit time for all the panels in Fig.\ref{fig:shpvssqr} was set to
$T_{bit}=2.0$.  We extend each transition to this time by appending
the appropriate interval of fixed-point current.  Since we normalized
time using $\omega_R= 2 \pi \text{f}_r$ this implies that we
communicate using NRZ bits that last about one third of the relaxation
oscillation period $1/\text{f}_r$ of the fixed point $(S_0,N_0)$.

We compare the bit errors using shaped currents to square-like current
modulation where, basically, the current follows the NRZ bits but with
finite rise times, to be slightly more realistic. The up and down
transitions of the square-like driving current shown in
Fig.\ref{fig:shpvssqr}d was achieved using `tanh' functions. The
maximum slope of the current equals the maximum slope of the shaped
current shown in Fig.\ref{fig:shpvssqr}a. Since the bit time is short
compared to the period of the laser's intrinsic relaxation
oscillations, the state-space trajectory due to the modulation of 50
NRZ-bits is complicated and exhibits large excursions (see
Fig.\ref{fig:shpvssqr}f). Since the system does not reach either one
of the two fixed points at the end of a bit-period, there is
considerable dynamic coupling from past to future and hence
inter-symbol interference. In Fig.\ref{fig:shpvssqr}e we show the
output corresponding to the drive current above
(Fig.\ref{fig:shpvssqr}d). It is apparent that even without channel
noise the receiver will not be able to infer without error the bit
sequence that was to be transmitted.

%
%
%
%
%
%

To quantify the possible improved communication performance we
calculated bit error rates (BER).  For illustration, a simple additive
white Gaussian noise corrupted a dispersionless channel.  The overall
conclusions drawn will apply to more realistic channel models, since
the performance enhancement is due to avoiding inter-symbol
interference in the laser and should be largely independent of the
characteristics of the communication channel.

\begin{figure}
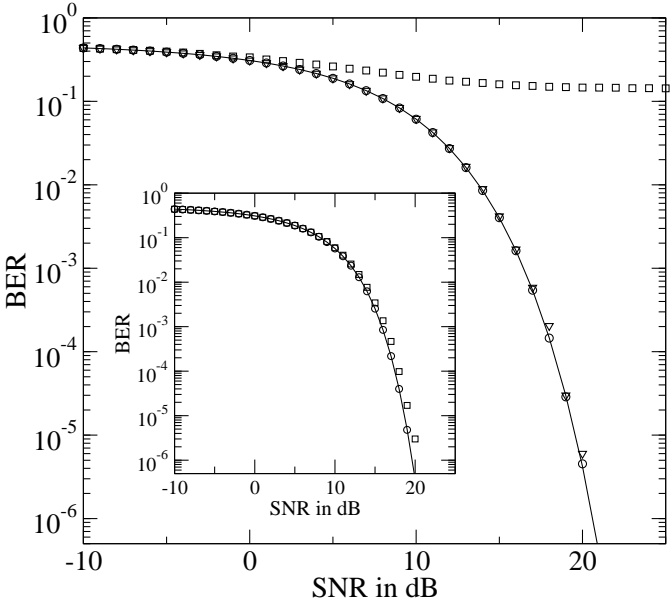

\includefigure{fig4_600dpi.eps}
\caption[]{ 
  The signal-to-noise ratio (SNR) versus the bit-error-rate (BER) is
  presented for the laser being driven by both shaped currents
  (circles and lines) and square-like currents (squares). The main
  panel shows high bit rate ($T_{bit}=2$) and the inset lower bit rate
  ($T_{bit}=8)$ communication. The symbols represent results obtained
  by numerical integration and the lines are theoretical curves. We
  display as triangles results where stimulated emission noise is
  simulated. }
\label{fig:BER}
\end{figure}

Figure~\ref{fig:BER} shows in the main panel the SNR versus the BER
for high bit rate ($T_{bit}=2$) square-like current modulation as
squares and high bit rate shaped current modulation as circles. Low
bit rate ($T_{bit}=8$) square-like current modulation (squares) and
low bit rate shaped current modulation (circles) is presented in the
inset of Fig.~\ref{fig:BER}.  These are results of numerical
experiments where one million random bits generated a corresponding
input current (either shaped or square-like) to (\ref{rate2}), which
was integrated. To the resulting laser output an appropriate amount of
channel noise was added.  The bits were decoded at the receiver by
integrating the received power over each bit time (assuming a perfect
clock) and by distinguishing zero and one bits through simple
thresholding. This way an estimate of the BER was obtained for each
noise level.

%
%
%

For the case of square-like current modulation the threshold used for
bit detection was determined numerically. It was optimized for each
noise level separately. Furthermore, due to the complex dynamics of
$s(\tau)$, it is not immediately obvious when to start the receiver
clock that determines the beginning and end of a bit (see
Fig.~\ref{fig:shpvssqr}e). Therefore the optimal starting time was
numerically estimated as well.

For a shaped current input, a simple theoretical expression for the
expected BER can be derived.  Consider transmission without channel
noise first. Without any noise the receiver detects four different bit
energies $E^i$ ($i\in\{1,2,3,4\}$) signifying one bits.%
\footnote{ \changed{ For the purpose of the numerical experiment we
    use $E=\int_0^{T_{bit}} s(t) dt$ for the bit energies because the
    photon density $s(t)$ is proportional to the laser output power.
  } } The largest energy $E^1$ corresponds to a one bit being both
preceded and succeeded by a one bit, the smallest energy $E^4$
corresponds to a one bit being both preceded and succeeded by a zero
bit, whereas $E^2$ and $E^3$ correspond to one bits that contain an up
or down transition \changed{ (see Fig.~\ref{fig:shpvssqr}).  }
Similarly, there are four energy levels ($E^j$ with $j\in\{5,6,7,8\}$)
representing zero bits.  The expected bit error rate is obtained by
evaluating the probability that the energy that results after adding
white Gaussian noise $\xi$ with variance $\sigma^2$ to a zero bit
energy level exceeds the threshold $E_{th}$ and by considering the
probability that $E^i+\xi<E_{th}$ for $i\in\{1,2,3,4\}$. Doing this
the theoretically expected BER can be written in terms of the
complementary error function: \be\label{theober} BER= \frac{1}{8}
\sum_{i=1}^{8} \frac{1}{2} \mbox{erfc} \left( \frac{|E^i -
    E_{th}|}{\sqrt{2} \sigma} \right).  \ee
The resulting curves are plotted as lines in Fig.\ref{fig:BER} where a
fixed threshold given by \be\label{thresh} E_{th}=\frac{1}{2} \left[
  \max_{j\in[5 \ldots 8]}( E^j) + \min_{i\in[1 \ldots 4]}( E^i)
\right] \ee was used.

Calculating the theoretically expected BER using the optimal threshold
value $E_{th}^{opt}(\sigma)$, which can be derived from
(\ref{theober}), and which is a function of the noise strength,
results in a curve that to the eye is indistinguishable from the one
displayed in Fig.\ref{fig:BER}.  For this reason we used the simpler
$E_{th}$ given by (\ref{thresh}) in the numerical experiment
determining the BER when using shaped modulation currents.

The following conclusion can be drawn from the results presented in
Fig.~\ref{fig:BER}.
As expected, using shaped current modulation there is a very good
agreement between the theoretical and numerical result for both high
and low bit rates, showing that all errors are due to Gaussian channel
noise.  In contrast, Fig.~\ref{fig:BER} shows that high bit rates
($T_{bit}=2$) for square-like current modulation lead essentially to a
communication breakdown. Even for very small noise levels (high SNR)
on average about 14 bits per 100 transmitted bits are wrongly decoded
by the receiver. These errors are caused by inter-symbol interference
due to the laser dynamics.  For lower bit rates, that is bit rates
smaller than $\omega_r / 2 \pi$, the bit errors attributable to the
laser dynamics become more and more negligible, as is indicated by the
inset of Fig.~\ref{fig:BER}, which shows that for lower bit rates the
performance enhancement due to the use of shaped input currents is
minor.

\subsection{Noise and Parameter Mismatch}

Semiconductor laser diodes are inherently noisy. To study the effects
of spontaneous emission noise on the proposed communication scheme we
conduct numerical experiments, where rate equations that include
appropriate stochastic terms (Gaussian white noise) are driven by the
shaped current, which, as discussed above, was calculated based on the
\changed{ deterministic model (\ref{rate2}).  The rate equation
  including spontaneous emission terms and a description of the
  integrator can be found in~\cite{article:Abarbanel,mythesis}.  }

For realistic values of the noise variance (same as
in~\cite{article:Abarbanel}) the performance advantage achieved by
modulating the laser with shaped input currents persists. Although
noise causes the state-space trajectory to deviate from the ideal
trajectory, the deviations remained for the bit sequences and noise
realization of our numerical experiments small enough to cause only
slight performance degradations. The BER performance calculated in one
such numerical experiment is shown as triangles in the main panel of
Fig.~\ref{fig:BER}. It is close to the theoretical curve calculated
for the noise free system and far better than square-like modulation
of a laser model free of spontaneous emission noise.

%
%
%
\begin{figure}
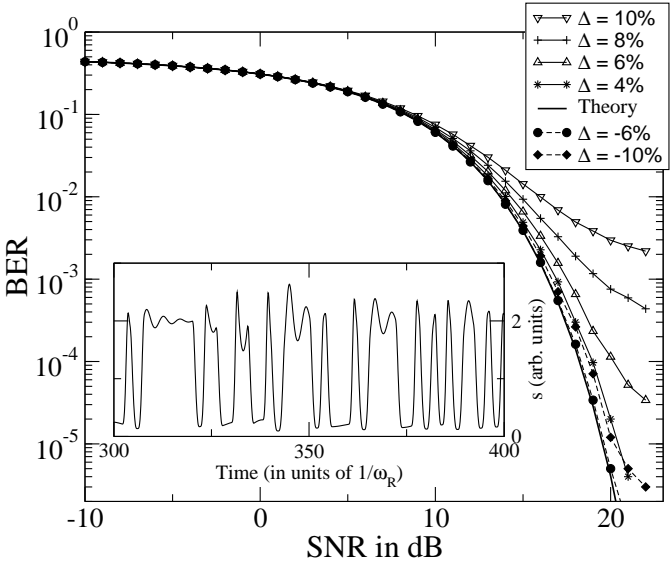

\includefigure{fig5_600dpi.eps}
\caption[]{\changed{
    BER performances for mismatches $\Delta=\frac{\gn'-\gn}{\gn'}$,
    where $\gn'$ is the estimated and $\gn$ the actual differential
    gain.  The inset shows a time trace of the photon density $s$ for
    $\Delta=-10\%$ at the output of the laser. The bits are clearly
    distinguishable.  } }
\label{fig:BERgn}
\end{figure}

The shaped current waveform that is used to achieve high bit rate
communications without inter-symbol interference was calculated based
on the model (\ref{rate2}), the parameters of which will be known only
approximately for a given laser diode. It is therefore important to
consider the effects of parameter mismatch between the model used to
generate the shaped driving current and the model used to numerically
evaluate the BER performance.

%

\changed{%
  Figure \ref{fig:BERgn} shows the degradation of the BER performance
  for differential gain mismatches, defined as
  $\Delta=\left(\gn'-\gn\right)/\gn'$. Here $\gn'$ denotes the
  estimated differential gain, the one used in the generation of the
  shaped drive current, and $\gn$ is the actual parameter, the one
  characterizing the system driven by the shaped currents.
  Overestimating of the differential gain ($\Delta>0$) severely
  degrades performance for mismatches $|\Delta|>5 \%$, whereas
  underestimation ($\Delta<0$) has a far less noticeable effect.
}%

\changed{%
  The reason is that overestimating the differential gain implies an
  overestimation of $\omega_R$, i.e. the response time of the actual
  laser is slower than was assumed. The } laser will have slower rise
and fall times and consequently a lower bit energy for a one bit that
is both preceded and succeeded by a zero bit. Although the mean bit
energy does not change significantly, the energy difference between
the maximum energy for a zero bit ($\max_{j\in[5 \ldots 8]}( E^j)$)
and the minimum energy for a one bit ($\min_{i\in[1 \ldots 4]}( E^i)$)
is decreased resulting in a higher BER.

A similar asymmetric performance degradation holds for mismatches in
the photon decay parameter $\gc$. Performance is most sensitive to
mismatches of $\gc$ and $\gn$ and least sensitive to mismatches in
$\gp$.

%
%

The unexpected asymmetry of the performance degradation with
estimation may suggests that, in practice, it may be wiser to solve
for state-space trajectories using slightly smaller values of
$\gamma_{c,n}$ than the best available experimental estimations of the
parameters.  This may reduce the risk to performance from fluctuations
in $\gamma_{c,n}$.

\section{Discussion}

%
%
%
\begin{figure}
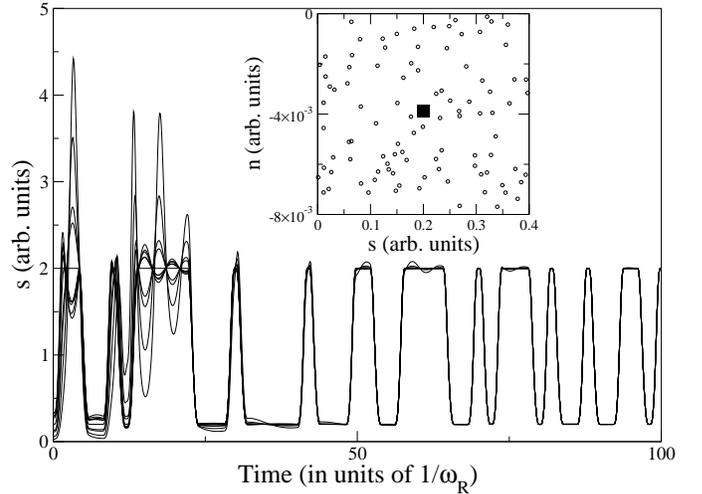

\includefigure{fig6_600dpi.eps}
\caption[]{ \changed {
    Evolution of the photon density s(t) in time for initial
    conditions chosen randomly in the neighborhood of the correct
    initial state (a subset of ten realizations is displayed, no noise
    and no mismatch).  }  Inset: Initial conditions in state-space
  (circles) and where the trajectories in state-space end up (squares)
  after driving currents for 49 random bits were applied.}
\label{fig:initial}
\end{figure}

We demonstrate a technique to control the laser's nonlinear internal
degrees of freedom to move along a specific state-space trajectory by
designing current input waveforms.  This can be used to optimize the
optical intensity waveforms emitted to fit certain desired
characteristics.
  
On the example of time-optimized NRZ bits, the possible BER
performance enhancement due to the elimination of inter-symbol
interference is demonstrated. We furthermore show that spontaneous
emission noise and parameter mismatches can be tolerated suggesting
that this scheme can be used in practice to enhance the communication
data rate when directly modulated lasers are used as transmitters.

Our numerical experiments show that including spontaneous emission
noise and small but finite parameter mismatches leads to trajectories
in state-space that are close to the ideal one. This suggest that the
trajectory we obtain in the optimization is robust with respect to
perturbations.  In numerical investigations it appears that the
state-space trajectory is also stable in the sense that, starting from
\changed{ slightly perturbed initial conditions, trajectories converge
  to the ideal one, } when the system is driven by the same bit
sequence.
The result of one such numerical experiment is shown in
Fig.~\ref{fig:initial}, \changed{ where initial relaxation
  oscillations soon disappear as the trajectories converge.  }

Theoretically, an open dynamical-systems question is whether there is
some kind of easily computable generalized stability property that
would quantify whether the controlled system is stable even under
arbitrary message driving.  It would be desirable to build this
property into the optimization target to ensure that the solutions
found are also stable.  In our model it was fortuitous that all
solutions were stable with the parameters we used, but this was not
guaranteed by the method, and at present the best check is empirical
integration under perturbations.  \changed{We also speculate that the
  relative advantage of dynamically-aware current-shaping over
  conventional direct modulation could be significantly larger if,
  instead of transmitting binary symbols (two energy levels), the
  system was controlled to a higher alphabet.  The net result of our
  pulse shaping is to radically reduce that part of variance which is
  due to dynamics in the optical intensities of the different received
  symbols.  If the fluctuations from inter-symbol dynamics are
  eliminated, leaving only detector noise and spontaneous emission,
  then it is conceivable that one could transmit with more than two
  distinct optical levels for the symbols, and still reliably decode
  them.  This would improve useful data rate versus binary
  transmission even without increasing the rate of symbols being
  transmitted.}

\changed {The key technological issue is whether the proposed
  direct-modulation scheme using shaped currents can be implemented at
  low cost.  Existing technology has been developed for clock
  equalization, generating fast fronts at gigahertz speeds which could
  subsequently be shaped using delay line techniques~\cite{dally}.
  Thus, current shaping is in principle electrically achievable.  We
  speculate that the laser and the current shaping electronics could
  be integrated on one chip so that the overall cost should be
  comparable to that of traditional single-mode-lasers.  We hope that
  the potential of current-shaping to mitigate problems that have
  plagued the direct-modulation approach to high-speed communication
  will stimulate further research and lead to practical applications.}

\normalem

\begin{biography}[{\includegraphics[width=1in,height=1.25in,clip,keepaspectratio]{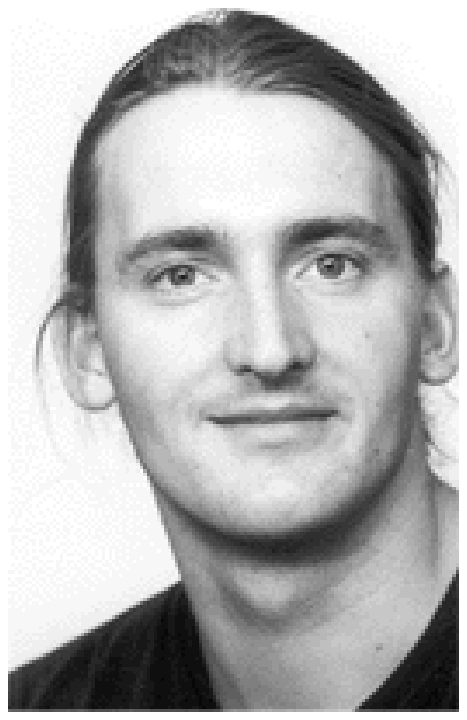}}]{Lucas Illing}
  was born in Halle, Germany, in 1973. He received his Vordiplom in
  1996 from the Humboldt University in Berlin, Germany and in 2002 the
  Ph.D. in Physics from the University of California, San Diego. He is
  a Research Assistant at Duke University.  His research is primarily
  concerned with chaos synchronization, control, and communication
  using chaos.
\end{biography}

\begin{biographynophoto}{Matthew B. Kennel}
  was born in Santa Monica, California in 1968.  He received the A.B.
  degree in physics in 1989 from Princeton University and in 1995 the
  Ph.D. in physics from the University of California, San Diego.
  After a Department of Energy postdoctoral fellowship at the Oak
  Ridge National Laboratory, he joined the Institute for Nonlinear
  Science at UCSD as an Assistant Research Scientist.  His research
  interests are in chaotic communication and the application of
  information theoretical algorithms to observed nonlinear dynamics.
\end{biographynophoto}

\end{document}